\documentclass[fleqn,12pt,twoside]{article}
\usepackage{espcrc1}
\usepackage{graphicx}
\usepackage[figuresright]{rotating}
\title{Recent Developments in Few-Nucleon Systems}

\author{A. Kievsky\address[INFN]{Istituto Nazionale di Fisica Nucleare, 
        and Physics Department, Universita' di Pisa, Via Buonarroti 2, 56100, Italy},
        M. Viviani\addressmark[INFN],
        L.E. Marcucci\addressmark[INFN]
        and
        S. Rosati\addressmark[INFN]}
       
\begin{document}

\maketitle

\def\bm{\boldmath}
\def\x{{\bf x}}
\def\y{{\bf y}}

\begin{abstract}
$N-d$ elastic scattering is studied at different energies using one
of the modern NN interactions, the
Argonne $v_{18}$ which explicitly includes the
magnetic moment interaction between two nucleons. This interaction, which 
has been often neglected in the description of the few-nucleon continuum,
produces sizable modifications in some elastic observables. Its
effects, as well as those produced by the Coulomb potential, are
analyzed as a function of energy. The magnetic moment interaction 
produces appreciable effects in $p-d$ scattering at low energies but
they are very small above $10$ MeV.
Above $65$ MeV Coulomb effects can be observed only in specific
observables as for example $T_{21}$.
\end{abstract}

\section{Introduction}

The three- and four-nucleon systems are excellent testing grounds
for the nuclear interaction. The last generation $NN$ interactions can
be used to calculate $3N$ and $4N$ bound and scattering states and,
from a comparison to the experimental data, important conclusions
about the capability of those interactions to reproduce the dynamics
can be extracted. It is widely accepted that the potential energy
in a few-nucleon system consists of a sum of the pairwise $NN$
interaction and a term including a pure three-nucleon interaction.
This last term is not very well known and, in general, its
strength is fixed so as to reproduce the experimental $A=3$ binding
energy. With the recent advances in the solution of the $3N$ and $4N$
continuum, the possibility of using scattering data to improve
our knowledge of the three-nucleon interaction is at present feasible.
Because of this, a correct treatment of the complete electromagnetic
interaction in the description of few-nucleon reactions is required. 
In fact, Coulomb effects and three-nucleon force effects are very often of 
the same size. Moreover, the magnetic moment (MM) interaction between two-nucleons 
can give sizable contributions to some polarization observables.

In the present contribution $N-d$ scattering is studied using
the Argonne $v_{18}$ (AV18) NN potential~\cite{av18}
plus the Urbana IX (UR) three-nucleon potential~\cite{urbana}
and electromagnetic terms as the Coulomb and magnetic moment
interactions. Different observables are considered such as the
differential cross section and vector and tensor analyzing powers. A
comparison to recent high-quality measurements is performed.

\section{The $N-d$ Transition Matrix including the MM Interaction}

Following the notation used in the determination of the
AV18 potential, 
all modern $NN$ potentials can be put in the general form

\begin{equation}
   v(NN)=v^{EM}(NN)+v^\pi(NN)+v^R(NN)\ .
\end{equation}

The short range part $v^R(NN)$ of these interactions
includes a certain number of parameters (around 40), 
which are determined by a fitting procedure to the $NN$ scattering data and
the deuteron binding energy (BE), whereas the long range part 
is represented by the one-pion-exchange potential $v^\pi(NN)$
and the electromagnetic potential $v^{EM}(NN)$. 

The AV18 potential includes the same $v^{EM}(NN)$ used in the Nijmegen 
partial-wave analysis except for short-range terms and finite size
corrections. The $v^{EM}(pp)$
consists of the one- and two-photon Coulomb terms plus the
Darwin-Foldy term, vacuum polarization and MM interactions.
The $v^{EM}(np)$ interaction includes a Coulomb term due to the neutron charge
distribution in addition to the MM interaction. Finally,
$v^{EM}(nn)$ is given by the MM interaction only. All
these terms take into account the finite size of the nucleon charge
distributions. Explicitly the long range part of 
the $NN$ magnetic moment interaction reads:

\begin{eqnarray}
v_{MM}(pp) & = & -{\alpha\over 4M^2_p}\left[\mu^2_p {S_{ij}\over r^3} +
 (8\mu_p-2) {{\bf L}\cdot {\bf S} \over r^3} \right]  
 \;\; , \label{mmpp} \\
v_{MM}(np) & = & -{\alpha\; \mu_n \over 4 M_n M_p}\left[ \mu_p {S_{ij} \over r^3}
 +{ 2 M_p \over M_{np}} {({\bf L}\cdot {\bf S} +{\bf L}\cdot{\bf A}) \over r^3} 
  \right] \;\; , \label{mmnp}  \\
v_{MM}(nn) & = & -{\alpha\; \mu_n^2 \over 4M^2_n} {S_{ij}\over r^3}  \;\; . 
 \label{mmnn}
\end{eqnarray}
$M_p$ ($M_n$) is the proton (neutron)
mass and $M_{np}$ is the $n-p$ reduced mass. The MM interaction presents
the usual $r^{-3}$ behavior and has an operatorial structure.
In the $np$ case, the last term includes 
an asymmetric force (proportional to 
${\bf A}=$(\mbox{\bm$\sigma$}$_i-$\mbox{\bm$\sigma$}$_j)/2$)
which mixes spin-singlet and spin-triplet states. 

For $A=2$ the contribution of the MM interaction to the
the scattering amplitude has been extensively studied~\cite{knutson78,stoks90}.
It has been shown that due to its $r^{-3}$ behavior the scattering
amplitude results in a slow convergent series whose leading terms can be summed
analytically. A similar analysis can be performed for
$N-d$ scattering. Accordingly, the $N-d$ transition matrix
$M$ is written as
\begin{eqnarray}
M^{SS'}_{\nu\nu'}(\theta)& = &f_c(\theta)\delta_{SS'}\delta{\nu\nu'}+
 f_{MM}(\theta)+
{\sqrt{4\pi}\over k}\sum_{LL'}^{L_{max}}\sum_{J}\sqrt{2 L+1}
(L0S\nu|J\nu)(L'M'S'\nu'|J\nu) 
\nonumber \\
&\times& \,\exp[i(\sigma_L+\sigma_{L'}-2\sigma_0)]\;
 {}^JT^{SS'}_{LL'} \; Y_{L'M'}(\theta,0) \;\; .
\label{tm}
\end{eqnarray}
The matrix $M$
is a $6\times6$ matrix corresponding to the projections of the
two possible couplings of the 
spin of the deuteron $S_d=1$ and the spin $1/2$ of the third particle 
to $S=1/2$ or $3/2$.
The quantum numbers $L,L'$ represent the relative orbital angular momentum between
the deuteron and the third particle and $J$ is the total angular momentum
of the three-nucleon scattering state. 
${}^JT^{SS'}_{TT'}$ are the $T$-matrix elements corresponding to a Hamiltonian
containing nuclear plus Coulomb plus MM interactions. 
In the above equation $f_c$ is the Coulomb amplitude:
\begin{equation}
f_c(\theta)=
 \sum_{L=0}^\infty(2L+1)({\rm e}^{2i\sigma_L}-1) P_L(\cos\theta)
 = -2i\eta\frac{{\rm e}^{2i\sigma_0}}{1-\cos\theta}
 {\rm e}^{-i\eta{\rm ln}(\frac{1-\cos\theta}{2})} \;\;\ ,
\end{equation}
and $f_{MM}$ is the amplitude generated by the inclusion of the MM
interaction
\begin{eqnarray}
 f_{MM}(\theta)& = & 
{\sqrt{4\pi}\over k}\sum^\infty_{L>L_{max}}\sum^\infty_{L'>L_{max}}\sum_{J}
    \sqrt{2 L+1} (L0S\nu|J\nu)(L'M'S'\nu'|J\nu) \nonumber \\
 && \times \exp[i(\sigma_L+\sigma_{L'}-2\sigma_0)]
 {}^JT^{SS'}_{LL'} \; Y_{L'M'}(\theta,0) \nonumber \\
 &=&f^{SS'}_{\mu\mu'}
 \left[\frac{\cos\theta+2{\rm e}^{-i\eta{\rm ln}(\frac{1-\cos\theta}{2})}-1}
 {\sin\theta} 
 -\sum_{L=1}^{L_{max}}\frac{(2L+1)}{L(L+1)} {\rm e}^{2i(\sigma_L-\sigma_0)}
   P^1_L(\cos\theta) \right] \;\;\ ,
\label{tm1}
\end{eqnarray}
with $f^{SS'}_{\mu\mu'}$ a $6\times 6$ matrix which take into account
the spin structure of the MM interaction~\cite{kievsky03}. In the
derivation of the above equation the following relation has been 
used~\cite{knutson78,knutsonp}:
\begin{equation}
 \sum_{L=1}^\infty\frac{(2L+1)}{L(L+1)} {\rm e}^{2i\sigma_L} P^1_L(\cos\theta)=
 \frac{{\rm e}^{2i\sigma_0}}{\sin\theta} 
 [\cos\theta+2{\rm e}^{-i\eta{\rm ln}(\frac{1-\cos\theta}{2})}-1] \;\; \ ,
\end{equation}
with $P^1_L$ a generalized Legendre polynomial. 
The $n-d$ transition matrix is recovered putting $f_c=\sigma_l=\eta = 0$.

If the MM interaction is not considered the sums over $L$, $L'$, $J$ in the last
term of eq.(\ref{tm}) converge very fast 
due to the finite range of the nuclear interactions. Typically in the low energy
region ($E_{lab}<50$ MeV) states with values of $L,L'>10$ can be safely
neglected. However
when the MM interaction is considered, an infinite number of terms contributes
to the construction of the scattering amplitude. In this case the sums on $L,L'$ 
have been divided in two parts depending on a certain value of $L_{max}$.
For $L,L' \le L_{max}$ the
$T$--matrix elements correspond to, and are obtained from,
a complete three-body description of the system. 
For $L,L' > L_{max}$ the centrifugal barrier is sufficiently
high to maintain the third particle far from the deuteron and the description
of the state can be performed as a two-body system. Moreover, in this regime
the nuclear interaction can be neglected and
the interaction between the incident nucleon and the deuteron 
can be considered only as electromagnetic.
The $T$-matrix elements corresponding to values of
$L,L' > L_{max}$ have been obtained in Born approximation as
\begin{equation}
  ^J\!T^{LL'}_{SS'}=-k ({2M_{nd}\over \hbar^2})
  <\Omega_{L'S'J}|v_{MM}(Nd)|\Omega_{LSJ}>  \;\; ,
\label{born1}
\end{equation}
with $M_{Nd}$ the $N-d$ reduced mass and 
$\Omega_{LSJ}=F_L(kr)[Y_L(\hat r)\otimes \chi_S]_{JJ_z}$
is the $N-d$ wave function. Here $F_L$ represents the regular
Coulomb (Bessel) function for the $p-d$ ($n-d$) system
and $k^2=(2M_{nd}/\hbar^2)E_{cm}$. 
In the above equation the MM interaction between a nucleon and
the deuteron appears explicitly and it is defined as:
\begin{eqnarray}
v_{MM}(nd)&=&-{\alpha\over r^3}[ {\mu_n\mu_d\over M_n M_d} S_{nd}^I
       +{\mu_n\over 2 M_n M_{nd}}
        ({\bf L}\cdot{\bf S}_{nd}+{\bf L}\cdot{\bf A}_{nd})]\;\; , \label{ndMM} \\
v_{MM}(pd)&=&-{\alpha\over r^3}[ {\mu_p\mu_d\over M_p M_d} S_{pd}^I
       +({\mu_p\over 2 M_p M_{pd}}-{1\over 4 M^2_p})
        ({\bf L}\cdot{\bf S}_{pd}+{\bf L}\cdot{\bf A}_{pd})\nonumber \\
    && +({\mu_d\over 2 M_d M_{pd}}-{1\over 4 M^2_d})
        ({\bf L}\cdot{\bf S}_{pd}-{\bf L}\cdot{\bf A}_{pd})
       -{Q_d\over 2} S^{II}_d]\;\; ,  \label{pdMM} \\
    S^I_{Nd} &=& 3({\bf S}_N\cdot{\hat r})({\bf S}_d\cdot{\hat r})
       -{\bf S}_N\cdot{\bf S}_d,\;\;\;\; N=n,p  \\
    S^{II}_d &=& 3({\bf S}_d\cdot{\hat r})^2 - 2  \;\; ,
\end{eqnarray}
where $M_d$ is the deuteron mass and
$Q_d$ is the deuteron quadrupole moment.
${\bf S}_{Nd}={\bf S}_N+{\bf S}_d$ whereas ${\bf A}_{Nd}={\bf S}_N-{\bf S}_d$.
The deuteron-nucleon distance is $r$ and $\hat r$ is the unitary vector giving 
their relative position.

The $T$-matrix elements of eq.(\ref{born1}) are used to determine
the $f_{MM}$ amplitude of eq.(\ref{tm1}). They
originate a partial wave series which convergences slowly. In fact,
after performing the spatial integrals in eq.(\ref{born1})
a term proportional to $\delta_{LL'}/[L(L+1)]$ appears that
has been summed analytically up to $\infty$ in eq.(\ref{tm1}).

\section{N-d observables including MM interaction}

Elastic observables for $N-d$ scattering can be calculated using
the transition matrix of eq.(\ref{tm}) using trace
operations~\cite{seyler}. The calculations presented here have
been performed using the complex form of the Kohn variational 
principle~\cite{kievsky97} after an expansion of the three-nucleon
scattering wave function in terms of the pair correlated hyperspherical
harmonic basis~\cite{kievsky93,kievsky94}. The method used to describe
$n-d$ as well as $p-d$ elastic scattering is given in Ref.~\cite{kievsky01}.
The $NN$ interaction we have used is the nuclear part of the AV18
potential with and without the MM interactions defined in
eqs.(\ref{mmpp}-\ref{mmnn}). The asymmetric force ${\bf L}\cdot{\bf A}$
in the $v_{MM}(np)$ interaction has been considered.

At energies below the deuteron breakup the contribution of the MM interaction is
expected to be appreciable.  The $n-d$ neutron
analyzing power $A_y$ has been recently measured at $E_n=1.2$ and $1.9$ 
MeV~\cite{werner2002}. At these low energies the nuclear part
of the transition matrix (last term of eq.(\ref{tm})) converges
already for $L_{max}=3$. Different theoretical curves have been computed
and compared to the experimental data in Fig.1 at $E_n=1.2$ MeV (left panel)
and at $E_n=1.9$ MeV (right panel). The solid lines correspond to calculations using
the AV18 potential and neglecting the MM interaction.
These calculations show the usual
underestimation that all modern $NN$ forces produces in the description of
$A_y$. When the MM interaction is taken into
account the analyzing powers calculated at both energies, using the last term of
eq.(\ref{tm}) up to $L_{max}=3$, are given by the dashed curves. We can
observe a very small influence of the MM interaction in the peak of $A_y$ 
with the tendency of slightly flattening the observable. 
However, this is an incomplete calculation since the inclusion of the
MM interaction requires an infinite number of partial waves in the 
calculation of the transition matrix. When the MM amplitude $f_{MM}$
is considered the observables are given by the 
dashed-dotted curves. It is interesting to notice the forward-angle
dip structure which already appears in $n-p$ scattering~\cite{stoks90}.
Only after summing the series up to $\infty$ it is possible to
reproduce this particular behavior which is a consequence of the term
proportional to $(\cos\theta + 1 ) / \sin\theta$. In fact this term
produces a divergence in the n-d differential cross section as
$\theta\rightarrow 0$. However this divergence appears at extreme
forward angles. In Fig.1 the dip structure is only partially shown since
$A_y=0$ at $\theta=0$.
We can conclude that the MM interaction
produce a pronounced modification of $A_y$ at forward angles but has a
very small effect near the peak. 
Besides the differential cross section and the vector analyzing power 
other elastic $n-d$ observables as the tensor analyzing powers suffer only
minor modifications when the MM interaction is included. The differences
are of the order of $1$\% or less.
\begin{figure}[h]
\includegraphics*[angle=-90,width=80mm]{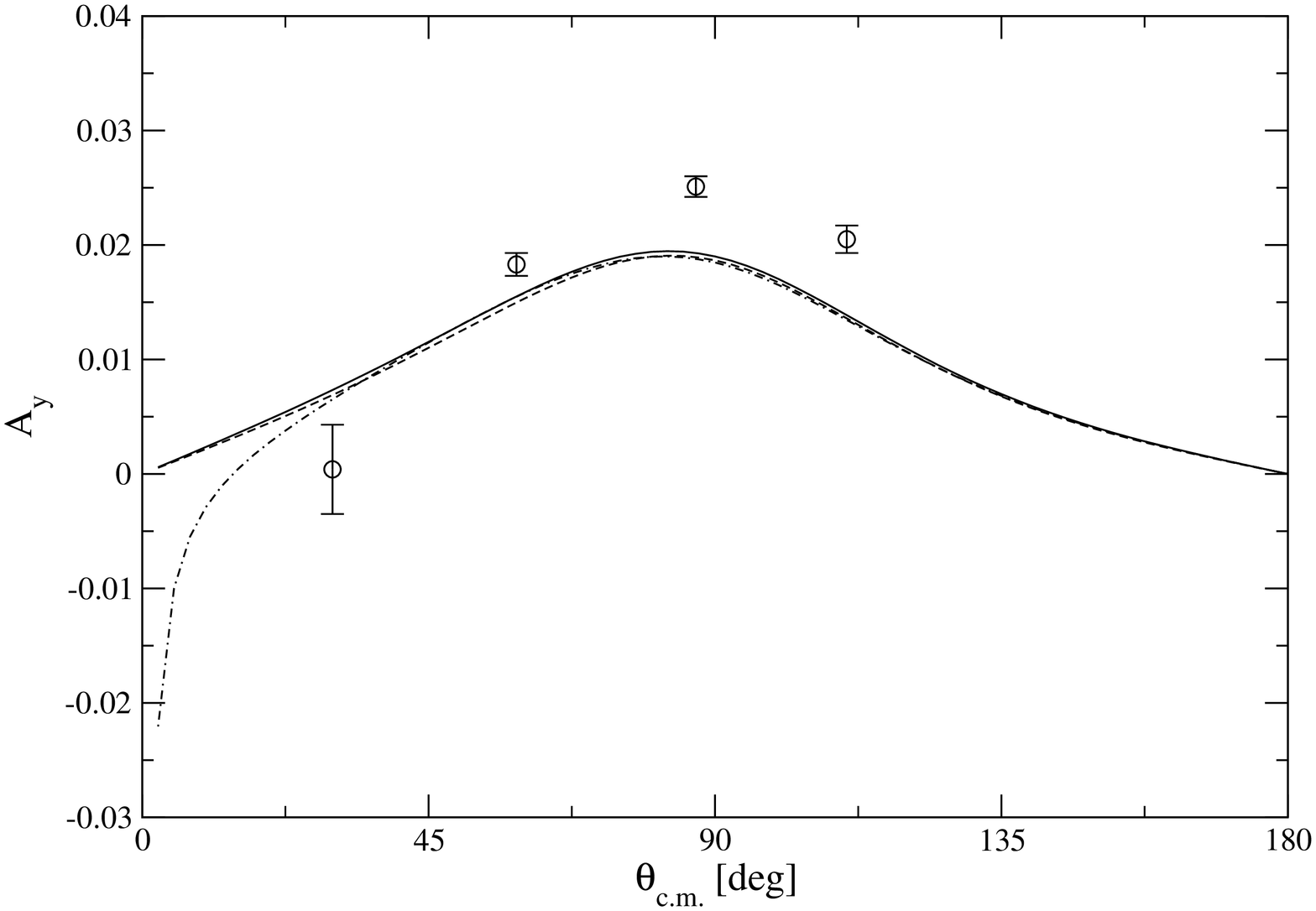}
\includegraphics*[angle=-90,width=80mm]{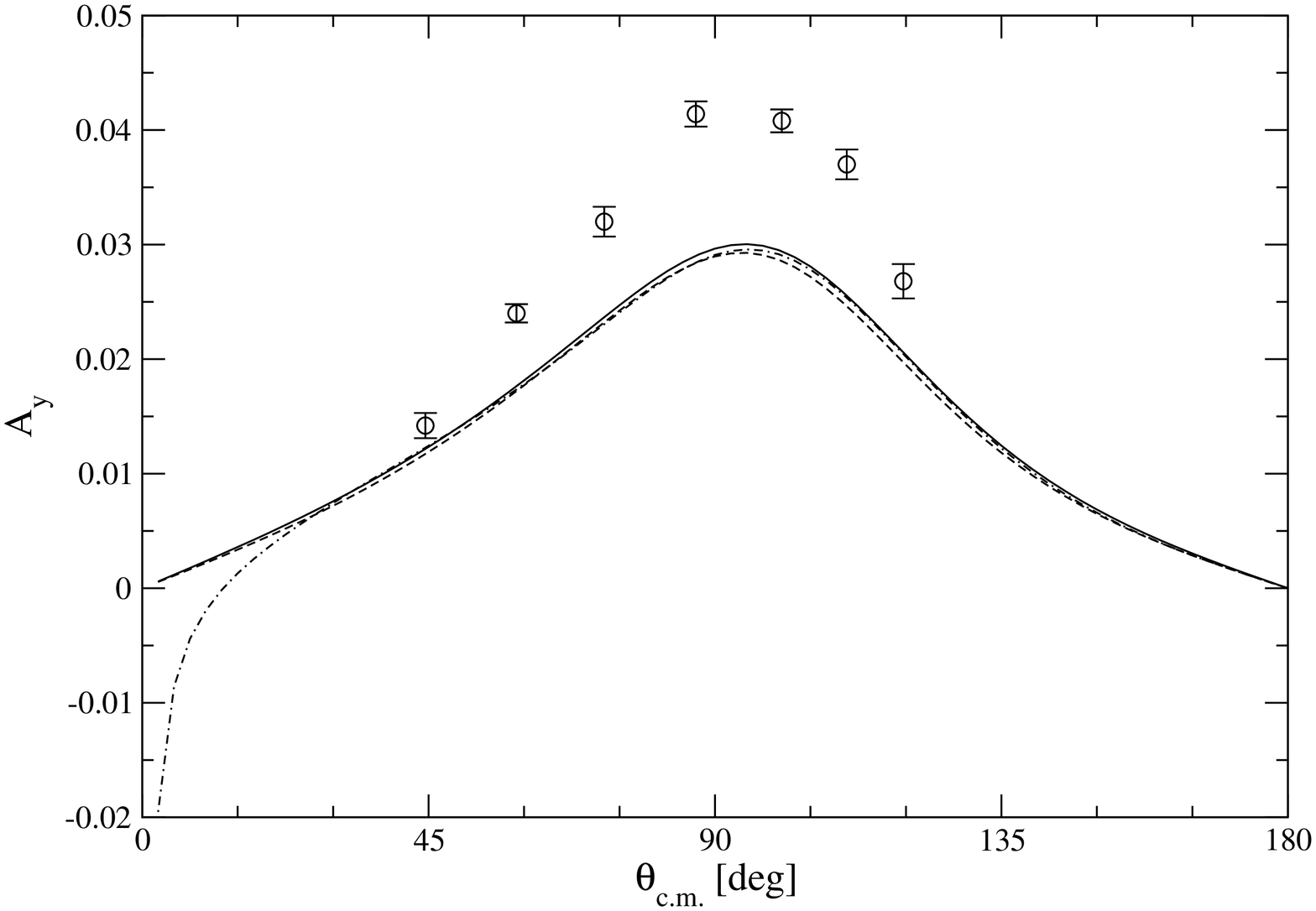}
\caption{The neutron $A_y$ at $E_n=1.2$ MeV and $1.9$ MeV. For explanation of
the curves see text. Experimental data are from Ref.~\cite{werner2002}}
\end{figure}

For $p-d$ scattering high precision data exist at low 
energies~\cite{brune,wood,knutson,sagara} for differential cross sections
and vector and tensor analyzing powers. Detailed comparisons to these
data has been performed in Refs.~\cite{brune,wood,kievwood,ktvr96,kvr95}
using AV18 with and without the inclusion of three-body forces. In these
studies the Coulomb interaction has been included whereas the MM interaction
has not. In order to evaluate the effects of the MM interaction on the
vector analyzing powers in presence
of the Coulomb field, in Fig.2 we show results at $E_p=1$ and $3$ MeV.
Three different calculations have been performed at both energies. The solid
line corresponds to the AV18 prediction neglecting the MM interaction. 
The partial-wave series has been summed
up to $L_{max}=4$ ($E_p=1$ MeV) and $L_{max}=6$ ($E_p=3$ MeV). 
The dashed line corresponds to the same calculation as before but the
$T$-matrix elements has been calculated using the AV18+MM potential.
The dashed-dotted line corresponds to the complete calculation
including the $f_{MM}$ amplitude. We see that the major effect
of the MM interaction is obtained around the peak and is appreciable
at both energies. There is also an improvement in the description of
the observable at forward angles. The observed modifications are due to the
interference between the Coulomb and the nuclear plus the MM interaction
and not due to high order terms as in the $n-d$ case. 
In fact high order terms are dominated by the Coulomb interaction
and the MM interaction gives a very small contribution.
As for $n-d$ scattering, the tensor analyzing
powers present very small modifications when the MM interaction is considered.
\begin{figure}[h]
\vspace{1cm}
\begin{center}
\includegraphics[width=14cm]{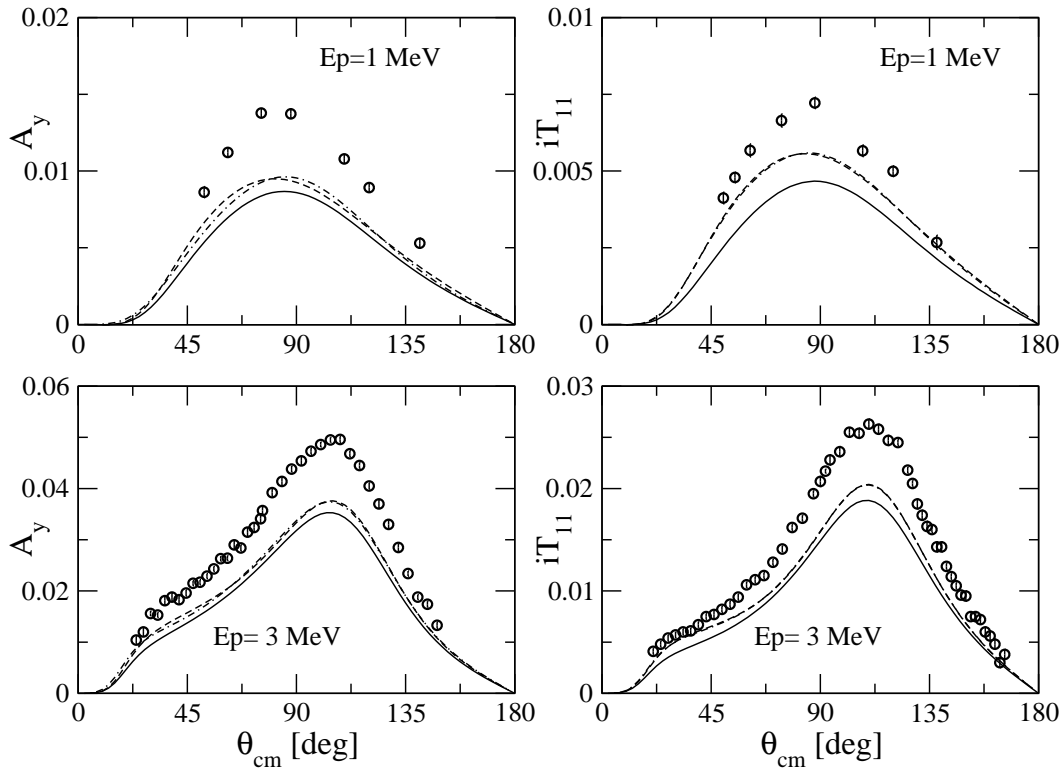}
\end{center}
\caption{The $p-d$ vector analyzing powers $A_y$ and $iT_{11}$. For explanation
of the curves see text. Experimental data are from Refs.~\cite{wood,sagara}
at $E_p=1$ MeV and $3$ MeV respectively.} 
\end{figure}

Increasing the energy the effect of the MM interaction on $A_y$ and $iT_{11}$
diminishes. Above $E_{lab}=10$ MeV calculations using AV18 or AV18+MM
give extremely close results~\cite{bled}. A different question is the importance
of Coulomb effects. 
In fact, up to $E_{lab}=30$ MeV we can observe appreciable 
differences in the description of $n-d$ and $p-d$ elastic scattering~\cite{kievsky01}.
Experimental data are not always conclusive since experiments with neutrons
have larger uncertainties than those performed with protons. Several times 
the description of
$p-d$ scattering has been performed using $n-d$ calculations, in particular 
at high energies~\cite{walter}. 
To make clear the importance of Coulomb effects as the energy of the collision
increases, in Figs.3,4 we show a comparison of 
$n-d$ and $p-d$ calculations at $E_{lab}=65$ MeV. 
In Fig.3 the
differential cross section and $A_y$ are shown. Three curves are displayed
corresponding to $p-d$ AV18 (solid line), $n-d$ AV18 (dashed line) and
$p-d$ AV18+UR (dotted line) and compared to experimental data. 
In Fig.4 the same calculations are shown
for $iT_{11}$ and the three tensor analyzing powers. We can observe
appreciable Coulomb effects in $T_{21}$ whereas three-nucleon interaction
effects can be observed in the minimum of the differential cross
section and in $T_{21}$ and $T_{22}$ as well.

\begin{figure}[htb]
\vspace{1cm}
\begin{center}
\includegraphics[width=14cm]{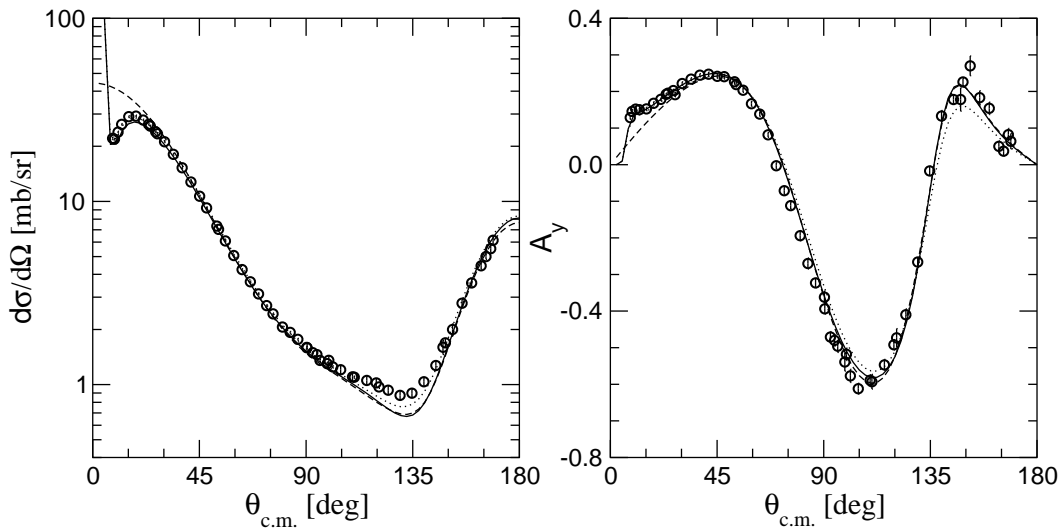}
\end{center}
\caption{ The differential cross section and $A_y$ at $E_{lab}=65$ MeV.
For explanation of the curves see text. 
Experimental data are from Ref.~\cite{shimizu}}
\end{figure}

The MM interaction has a different effect in $n-d$ or $p-d$
vector analyzing powers. One reason is the different sign of the neutron 
and proton magnetic moments. The other reason is the interference with the 
Coulomb field. However the MM interaction does not help for a better description
of the neutron $A_y$. On the contrary there is an appreciable improvement in the
proton $A_y$ as well as in $iT_{11}$, in particular at very low energies. 
Hence we can study the differences
between the experimental data and the calculations at the peak and see if the
inclusion of MM interaction helps to clarify a different behavior observed 
for $n-d$ and $p-d$. In Fig.5 we show the relative difference 
$[A_y(exp.)-A_y(th.)]/A_y(exp.)$ at the peak for $n-d$ and for $p-d$ scattering. 
In this
last case the AV18 and AV18+MM results have been considered. For $n-d$ both
results are extremely close at the peak and only the AV18 result has been
considered. We see that without the inclusion of the MM interaction
the underestimation of the proton $A_y$ is much more pronounced than the
neutron $A_y$. When the MM interaction is considered both, the $n-d$ and $p-d$
$A_y$ are predicted with a reduction of similar size (around $25$\%) 
in all the energy interval below $16$ MeV. 
Above $16$ MeV the differences on the peak between theory and experiment diminish
and practically disappear at $30$ MeV. In Fig.5 we see that at $18$ MeV the
difference has been reduced to $20$\%.

\section{conclusions}

We have shown that it is possible to perform detailed calculations 
to describe $N-d$ elastic scattering using a potential including the
nuclear plus Coulomb plus MM interactions. 
Though the strength of the MM interaction is small compared to the nuclear one,
its long range produces appreciable effects in some observables
as the $n-d$ differential cross section and vector analyzing powers.
For $p-d$ scattering we observed an increase
in vector analyzing powers at low energies. The importance of Coulomb
effects has been analyzed in Ref.~\cite{kievsky01} below $30$ MeV. Here we
have shown that at $65$ MeV Coulomb effects are very reduced in most
of the elastic observables with the exception of $T_{21}$ where an
appreciable effect has been observed in the peak.

In the low energy regime the electromagnetic potential produces
modifications in the observables that
are of similar size to those ones produced by the most frequently used
three--nucleon interactions, as the UR potential. Therefore when the three-nucleon
continuum would be used to study the structure of the three--nucleon
interaction, the complete electromagnetic NN potential should be
taken into account.

\vspace{1cm}

\begin{figure}[h]
\begin{center}
\includegraphics[width=15cm]{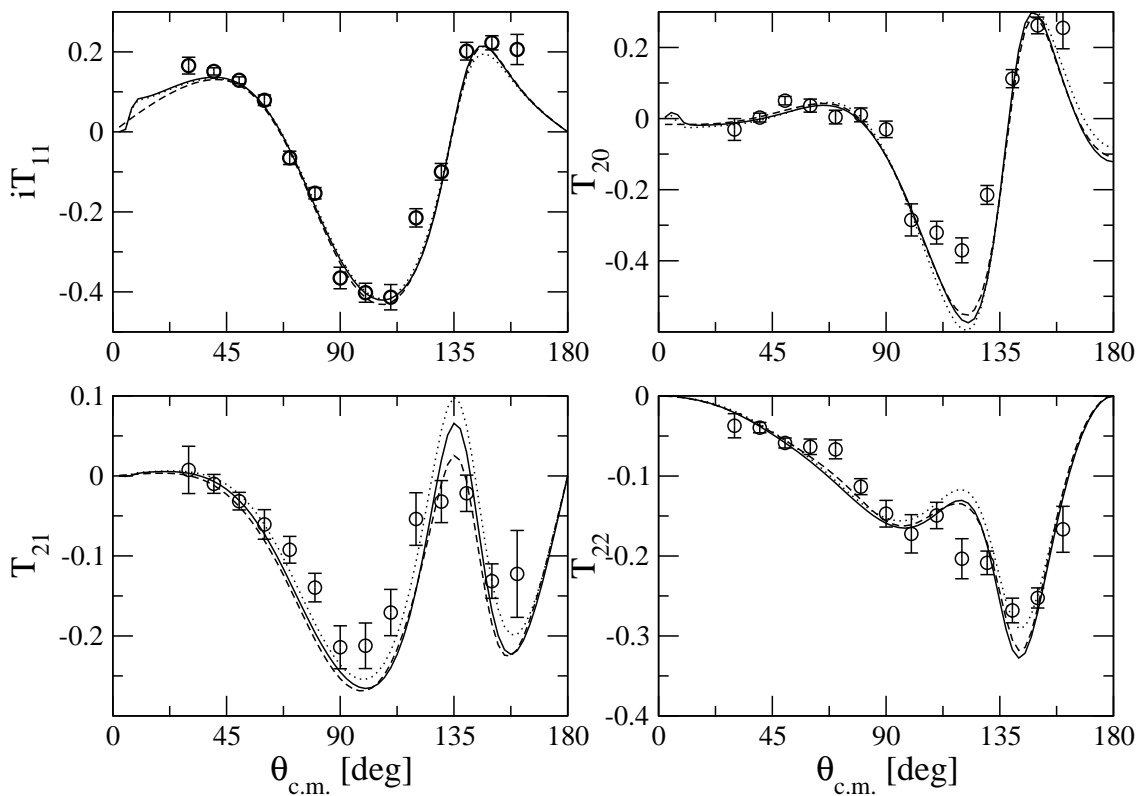}
\end{center}
\caption{The deuteron analyzing power $iT_{11}$ and the tensor analyzing
powers $T_{20},T_{21},T_{22}$ at $E_{lab}=65$ MeV.
For explanation of the curves see text. 
Experimental data are from Ref.~\cite{witala}}
\end{figure}
\begin{figure}[h]
\begin{center}
\includegraphics[width=95mm]{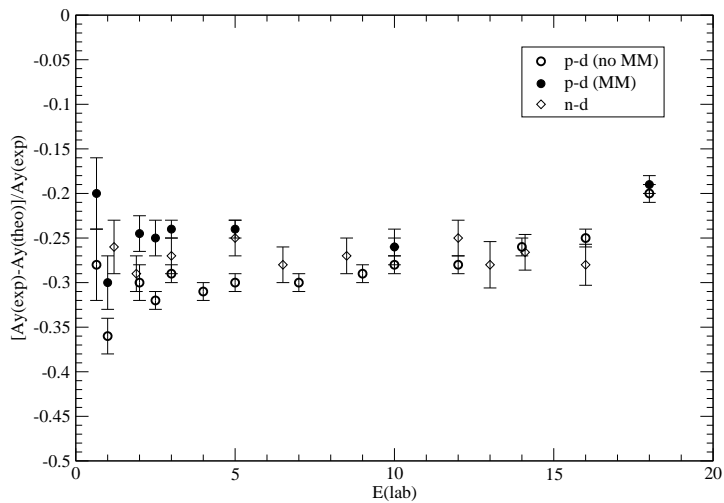}
\caption{Relative difference at the peak of $A_y$ as a function of
energy.}
\end{center}
\end{figure}

\end{document}